\documentclass[aps,prl,twocolumn,superscriptaddress,longbibliography]{revtex4-2}
\pdfoutput=1
\usepackage{graphicx}
\usepackage{amssymb,amsmath}
\usepackage{bm,dsfont}
\usepackage{dcolumn}
\usepackage{float}
\usepackage[OT1]{fontenc} 
\usepackage{url}
\usepackage{mathrsfs}
\usepackage{slashed,comment}
\usepackage{color}
\usepackage{verbatim}
\usepackage{enumitem}
\usepackage{soul,physics}
\usepackage[driverfallback=dvipdfm]{hyperref}
\hypersetup{pdfpagemode=FullScreen,colorlinks=true,breaklinks,urlcolor=blue,linkcolor=blue,citecolor=blue}

\usepackage{subfigure}
\usepackage{amssymb}
\usepackage{bm}
\usepackage{graphicx}
\usepackage{amsmath,amssymb}

\usepackage{pdfpages} 
\usepackage{pgffor} 

\makeatletter
\AtBeginDocument{\let\LS@rot\@undefined}
\makeatother

\def\supplementfilename{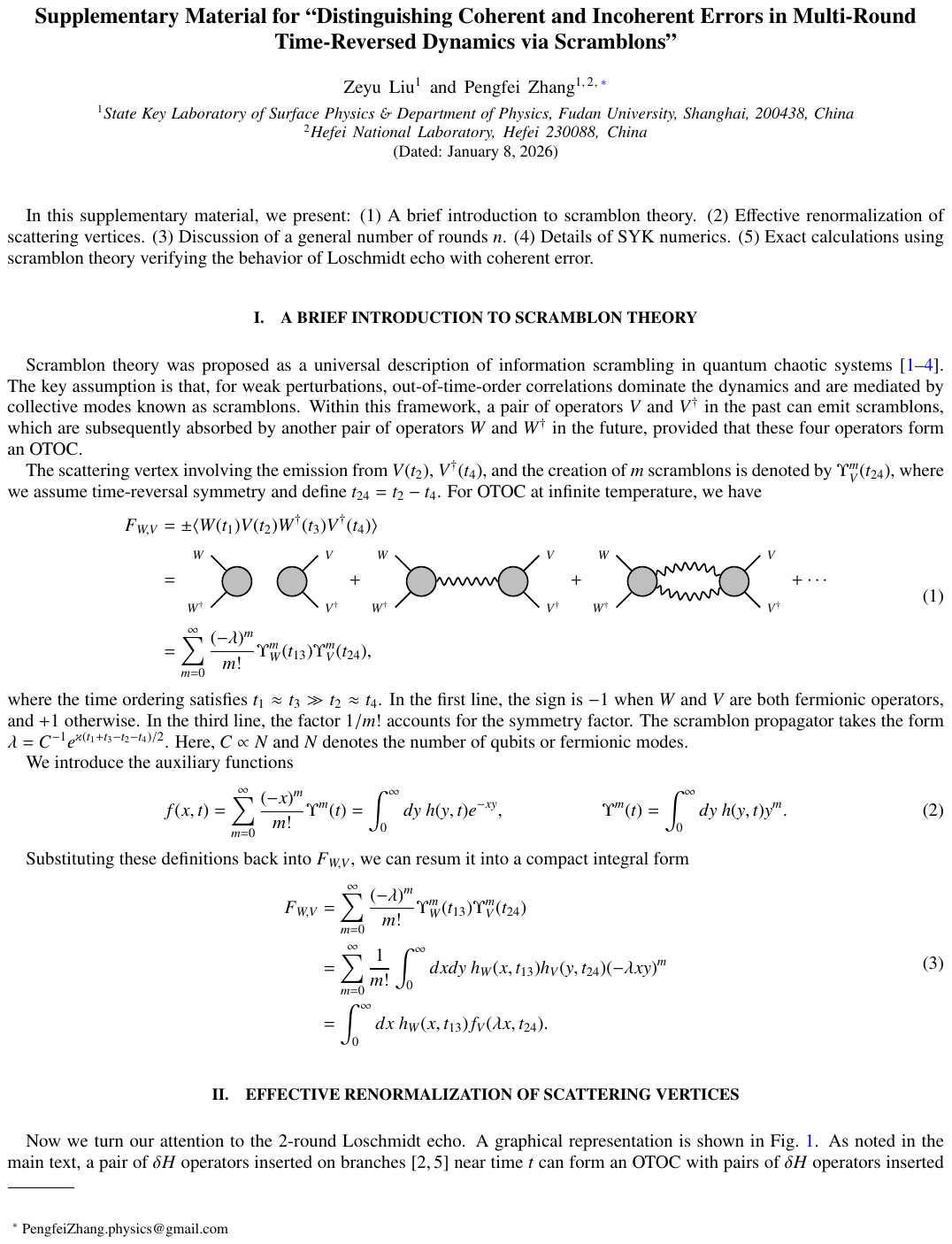}

\pdfximage{\supplementfilename}
\def\numbersupplementpages{\the\pdflastximagepages}

\newif\ifarXiv
\arXivtrue 
\usepackage{tikz,fp}
\usepackage{tikz-cd}
\usetikzlibrary{arrows}
\usetikzlibrary{intersections}
\usetikzlibrary{shapes.geometric}
\usetikzlibrary{decorations.pathmorphing, patterns,shapes,fixedpointarithmetic}
\usetikzlibrary{decorations.markings}

\pgfdeclarepatternformonly{south west lines}{\pgfqpoint{-0pt}{-0pt}}{\pgfqpoint{3pt}{3pt}}{\pgfqpoint{3pt}{3pt}}{
    \pgfsetlinewidth{0.4pt}
    \pgfpathmoveto{\pgfqpoint{0pt}{0pt}}
    \pgfpathlineto{\pgfqpoint{3pt}{3pt}}
    \pgfpathmoveto{\pgfqpoint{2.8pt}{-.2pt}}
    \pgfpathlineto{\pgfqpoint{3.2pt}{.2pt}}
    \pgfpathmoveto{\pgfqpoint{-.2pt}{2.8pt}}
    \pgfpathlineto{\pgfqpoint{.2pt}{3.2pt}}
    \pgfusepath{stroke}}

\tikzset{
    mid arrow/.style={postaction={decorate,decoration={
                markings,
                mark=at position .575 with {\arrow{stealth}}
    }}},
    near arrow/.style={postaction={decorate,decoration={
                markings,
                mark=at position .275 with {\arrow{stealth}}
    }}},
    far arrow/.style={postaction={decorate,decoration={
                markings,
                mark=at position .800 with {\arrow{stealth}}
    }}},
    snake arrow/.style={fixed point arithmetic, decorate, decoration={snake,amplitude=2pt, segment length=11pt},postaction={decoration={markings,mark=at position 0.625 with {\arrow{stealth}}},decorate}},
}
\tikzset{
  baseline = -0.5ex,
  wavy/.style = {
    thick,
    decorate,
    decoration={snake,amplitude=2pt,segment length=5pt}},
  sdot/.style = {
    circle,
    draw=none,
    fill=black,
    minimum size=2.5pt,
    inner sep=0pt},
  bdot/.style = {
    circle,
    draw=none,
    fill=black,
    minimum size=4pt,
    inner sep=0pt},
  svertex/.style = {
    circle,
    draw=black,
    thick,
    fill=lightgray,
    minimum size=8pt,
    inner sep=1pt},
  bvertex/.style = {
    circle,
    draw=black,
    thick,
    fill=lightgray,
    minimum size=24pt},
  bvertexsmall/.style = {
    circle,
    draw=black,
    thick,
    fill=lightgray,
    minimum size=7pt},
  bvertexnormal/.style = {
    circle,
    draw=black,
    thick,
    fill=lightgray,
    minimum size=16pt},
  dvertex/.style = {
    circle,
    draw=black,
    thick,
    fill=gray,
    minimum size=25pt}}

\begin{document}
	
	\title{Distinguishing Coherent and Incoherent Errors in Multi-Round \\Time-Reversed Dynamics via Scramblons }
	
    \author{Zeyu Liu}
    \affiliation{State Key Laboratory of Surface Physics \& Department of Physics, Fudan University, Shanghai, 200438, China}

    \author{Pengfei Zhang}
    \thanks{PengfeiZhang.physics@gmail.com}
    \affiliation{State Key Laboratory of Surface Physics \& Department of Physics, Fudan University, Shanghai, 200438, China}
    \affiliation{Hefei National Laboratory, Hefei 230088, China}

	\date{\today}
	\begin{abstract}
    Despite the rapid development of quantum science and technology, errors are inevitable and play a crucial role in quantum simulation and quantum computation. In quantum chaotic systems, coherent errors arising from imperfect Hamiltonian control and incoherent errors induced by coupling to the environment are both exponentially amplified during time evolution due to information scrambling. A fundamental question is how these two classes of errors imprint distinct signatures on the emergent irreversibility of many-body dynamics. In this Letter, we address this question by investigating multi-round time-reversed dynamics in the presence of both coherent and incoherent errors. By applying scramblon theory, we obtain closed-form expressions for the Loschmidt echo over different rounds of time-reversed evolution. For incoherent errors, the error accumulates linearly with the number of rounds, whereas coherent errors exhibit a crossover from quadratic to linear accumulation. These predictions are explicitly verified using the solvable Sachdev-Ye-Kitaev model. Our results provide a theoretical foundation for characterizing and calibrating coherent and incoherent errors in reversed dynamics, with particular relevance to nuclear magnetic resonance systems.
	\end{abstract}
	
	\maketitle

	\emph{ \color{blue}Introduction.--} The rapid development of quantum science and technology poses new challenges for the precise control of quantum many-body dynamics. In particular, the ability to reverse quantum many-body evolution is crucial both for experimentally probing quantum information dynamics and for implementing practical quantum algorithms \cite{KaiserLocalizationdelocalizationTransitionDynamics2015,Li:2016xhw,CappellaroExploringLocalizationNuclear2018a,CappellaroEmergentPrethermalizationSignatures2019,LiExperimentalObservationEquilibrium2020,2019arXiv190206628S,2021PhRvA.104a2402D,2022PhRvA.105e2232S,DuEmergentUniversalQuench2024a,Garttner:2016mqj,2019Natur.567...61L,ReyUnifyingScramblingThermalization2019b,LinkeExperimentalMeasurementOutofTimeOrdered2022,ChenInformationScramblingQuantum2021,DuanInformationScramblingDynamics2022,OliverProbingQuantumInformation2022,ZhaoProbingOperatorSpreading2022,ZobristConstructiveInterferenceEdge2025,VuleticTimereversalbasedQuantumMetrology2022,VuleticImprovingMetrologyQuantum2023,ChinQuantumSimulationUnruh2019,ThomasEnergyResolvedInformationScrambling2021,LiObservationQuantumInformation2024a,YouObservationAnomalousInformation2024a,WeidemullerTimereversalDipolarQuantum2024,gao2025signal}. An intrinsic obstacle to realizing such time-reversed dynamics with high precision is the inevitable presence of errors during evolution. There are two distinct sources of such errors. First, weak coupling of the system to its environment leads to decoherence, which is referred to as an incoherent error. Second, the Hamiltonian governing the backward evolution may exhibit small deviations from that of the forward evolution, resulting in coherent errors. Despite their different physical origins, both types of errors lead to similar phenomenology: their effects are exponentially amplified in time in chaotic systems, a universal feature of quantum many-body chaos, or quantum butterfly effect \cite{shenkerBlackHolesButterfly2014,Roberts:2014isa,Maldacena_2016,Shenker:2014cwa,kitaev2015simple}. This makes it difficult to directly distinguish coherent and incoherent errors in these time-reversal protocols. 

  \begin{figure}[t]
    \centering
    \includegraphics[width=0.80\linewidth]{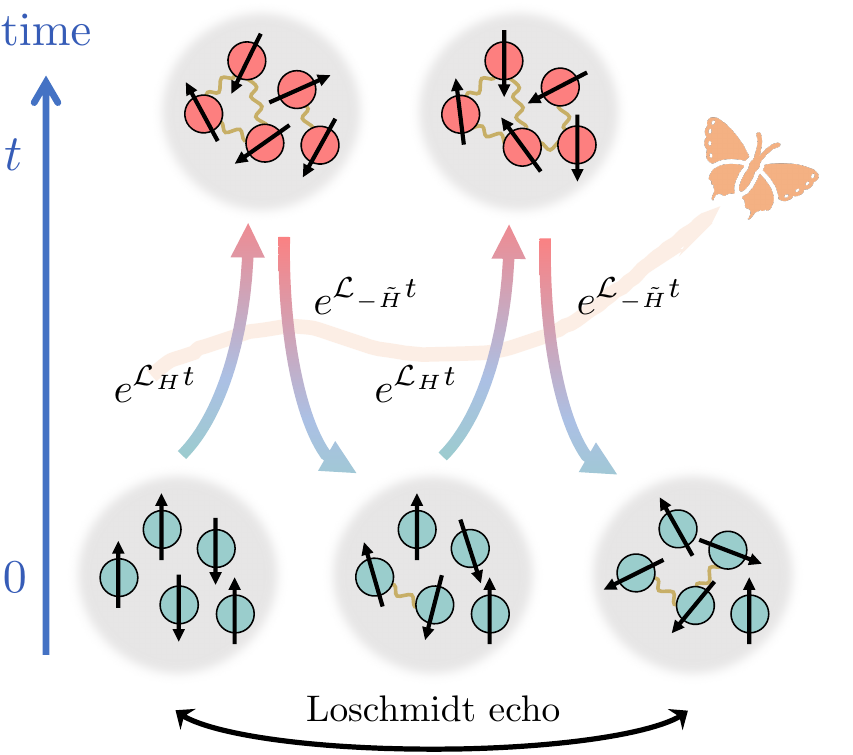}
    \caption{Schematic of the multi-round time-reversed dynamics, illustrated using the example of two rounds. Taking decoherence effects into account, the forward evolution is described by a Lindbladian superoperator $\mathcal{L}_H$ with Hamiltonian $H$. The Hamiltonian in the backward evolution, denoted by $\tilde{H}$, exhibits small deviations from $H$. These coherent and incoherent errors prevent perfect time reversal, with their effects being amplified by the quantum butterfly effect. }
    \label{fig:schemticas}
  \end{figure}

    Therefore, to explore the distinct signatures of coherent and incoherent errors in time-reversed dynamics, we consider a multi-round time-reversed protocol, as illustrated in Fig. \ref{fig:schemticas}. Starting from a high-temperature initial state, the system undergoes multiple forward and backward evolutions, with imperfections arising from either coherent or incoherent errors. Similar protocols have been adopted experimentally to measure high-order out-of-time-order correlators (OTOCs) \cite{Abanin:2025rbz}, which provide fine-grained measures of information scrambling \cite{Vardhan:2025rky}. To develop a general theory of the resulting dynamics for systems with all-to-all connectivity, we employ scramblon theory, which was proposed as a universal description of information scrambling in quantum chaotic systems \cite{Kitaev:2017awl,Gu:2018jsv,Zhang:2020jhn,Gu:2021xaj,Stanford:2021bhl,Zhang:2022knu,Zhang:2022fma,Liu:2023lyu,Stanford:2023npy,Zhang:2024vsa,Zhang:2025ckq,Chen:2024imd,Li:2025civ,Perugu:2025vty}. The key assumption is that, for weak perturbations, out-of-time-order correlations dominate the dynamics and are mediated by collective modes known as scramblons. Recent experiments have also validated scramblon theory in realistic solid-state nuclear magnetic resonance (NMR) systems using adamantane powder \cite{Li:2025civ}.

    We obtain closed-form results for the multi-round Loschmidt echo in time-reversed dynamics, which clearly elucidate the qualitative distinction between coherent and incoherent error scenarios. The effects of incoherent errors accumulate linearly with the number of rounds, reflecting the absence of inter-round correlations. In contrast, in the short-time regime, coherent errors exhibit a quadratic dependence on the number of rounds due to constructive inter-round interference, followed by a crossover to linear accumulation in the late-time regime. The crossover timescale depends logarithmically on the error magnitude, analogous to the logarithmic scaling of the scrambling time \cite{Maldacena:2016hyu}. Our predictions are explicitly demonstrated using the solvable Sachdev-Ye-Kitaev (SYK) model \cite{Sachdev:1992fk,kitaev2015simple,Maldacena:2016hyu,Kitaev:2017awl,Chowdhury:2021qpy} and are of direct experimental relevance to quantum platforms, particularly NMR systems \cite{NMR_UniversalDecay2008,NMR_UniversalDecay2011,NMR_UniversalDecay2012,KaiserLocalizationdelocalizationTransitionDynamics2015,NMR_TimeCrystal2018,CappellaroExploringLocalizationNuclear2018a,NMR_transport2011,Li:2016xhw,CappellaroEmergentPrethermalizationSignatures2019,2019arXiv190206628S,NMR_prethermal2021,NMR_hydrodynamics2023,DuEmergentUniversalQuench2024a,Li:2025civ}.

    \emph{ \color{blue}Setup.--} We now describe the details of the multi-round time-reversed dynamics for chaotic quantum many-body systems illustrated in Fig.~\ref{fig:schemticas}. We initialize the system in a high-temperature state $\rho_0 = e^{-\beta O}/\text{tr}[e^{-\beta O}] \propto (\mathds{1}-\beta O)$, where the inverse temperature $\beta \ll 1$. For nuclear magnetic resonance systems, the operator $O$ corresponds to the total spin along the magnetic field \cite{Li:2025civ}. The system evolves under a chaotic Hamiltonian $H$ with all-to-all interactions, together with decoherence arising from coupling to the environment. The resulting dynamics is described by the Lindblad master equation $\partial_t \rho=\mathcal{L}_H[\rho]$, with the Lindbladian superoperator
    \begin{equation}
    \mathcal{L}_H[\rho]= -i[H,\rho]+\sum_k \left(L_k \rho L_k^\dagger-\frac{1}{2}\{L_k^\dagger L_k,\rho\}\right).
    \end{equation}
    Here, $L_k$ are usually referred to as jump operators. The density matrix at time $t$ is therefore given by $\rho(t) = e^{\mathcal{L}_{H} t}[\rho_{0}]$. Next, we attempt to reverse the chaotic dynamics through control of the evolution Hamiltonian using techniques such as Floquet engineering \cite{Geier_Floquet2021,Martin_Thermalization2023,CappellaroEmergentPrethermalizationSignatures2019,NMR_prethermal2021}. In practice, this enables evolution under an effective Hamiltonian $-\tilde{H} = -H - \delta H$, where $\delta H$ denotes a small imperfection, corresponding to the coherent error. After backward evolution for a time $t$, the resulting density matrix becomes $\rho_1 = e^{\mathcal{L}_{-\tilde{H}} t} e^{\mathcal{L}_{H} t}[\rho_0]$, completing one round of time-reversed dynamics. In the multi-round protocol, this procedure is repeated for $n$ rounds, leading to $\rho_n=\left(e^{\mathcal{L}_{-\tilde{H}} t} e^{\mathcal{L}_{H} t}\right)^n [\rho_0]$. Finally, the measurement of operator $O$ is performed, which gives 
    \begin{equation}
    \langle O\rangle \propto \text{tr}\left(O\left(e^{\mathcal{L}_{-\tilde{H}} t} e^{\mathcal{L}_{H} t}\right)^n[O]\right)\equiv \text{tr}[O^2]F_n(t).
    \end{equation}
    Here, $F_n(t)$ is known as the Loschmidt echo \cite{Gorin:2006hhs,Quan:2006,Hasegawa:2021jcu,Jafari:2017kia}. This quantity probes imperfections in the reversed dynamics: in the absence of errors, $L_k = 0$ and $\delta H = 0$, the forward and backward evolutions cancel exactly, yielding $F_n = 1$. Our aim is to understand how coherent and incoherent errors manifest themselves in the decay of $F_n(t)$. For later convenience, we fix the normalization $\text{tr}[O^2]/\mathcal{D} = 1$, where $\mathcal{D}$ is the Hilbert space dimension.

  \begin{figure}[t]
    \centering
    \includegraphics[width=0.99\linewidth]{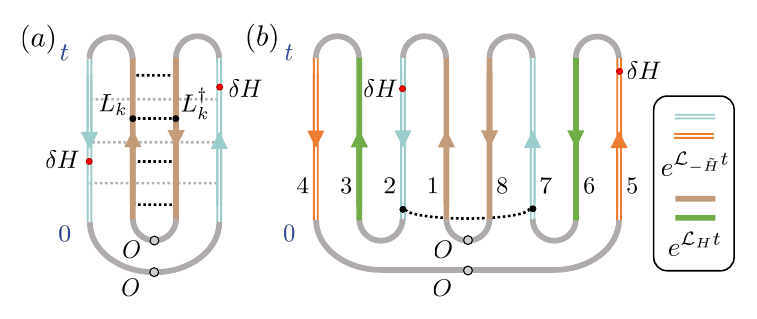}
    \caption{Graphical representation of the Loschmidt echo $F_n(t)$ for (a) $n=1$ and (b) $n=2$. Branches with the same color originate from the same Lindbladian evolution. Solid lines correspond to evolution governed by $\mathcal{L}_H$, while dotted lines indicate the Lindblad term. Double lines denote evolution in the presence of coherent errors, where insertions of $\delta H$, marked by red dots, may occur. In panel (b), all branches are labeled from 1 to 8. }
    \label{fig:contour}
  \end{figure}

    \emph{ \color{blue}Perturbation Theory.--} To understand the emergence of out-of-time-order (OTO) correlations in the Loschmidt echo, we first examine the single-round case. A graphical representation of the Loschmidt echo $F_1(t)$ is shown in FIG.~\ref{fig:contour}(a), with details provided in the figure caption. The special case involving only coherent errors was investigated in Refs.~\cite{Li:2025civ,LiTenn:2025gea}, whereas the case with only incoherent errors is closely related to the problem of operator growth in open systems \cite{Zhang:2019fcy,Almheiri:2019jqq,Weinstein:2022yce,Schuster:2022bot,Zhang:2022knu,Bhattacharya:2022gbz,PhysRevResearch.5.033085,Bhattacharjee:2022lzy,Bhattacharjee:2023uwx,Zhang:2024vsa,PhysRevD.110.086010,Garcia-Garcia:2024mdv,Jiang:2025lmn}. To the second order in jump operator $L_k$ and $\delta H$, the Loschmidt echo receives the contribution from 
    \begin{equation}\label{eqn:F1}
    \begin{aligned}
    F_1(t) \approx&1- 2\int_0^t dt'~\sum_k\langle OL_k^\dagger(t') [L_k(t'),O]\rangle\\&-\int_0^t dt'dt''~\frac{1}{2}\langle [O,\delta H(t')][\delta H(t''),O]\rangle.
    \end{aligned}
    \end{equation}
    Here, the expectation value is taken over the infinite-temperature ensemble $\mathds{1}/\mathcal{D}$, and operators are evolved under the errorless Hamiltonian, $M(t)=e^{iHt}Me^{-iHt}$. The factor of $2$ arises because incoherent errors $L_k$ can occur in both the forward and backward evolutions, whereas the coherent error $\delta H$ appears only in the backward evolution. Eq.~\eqref{eqn:F1} shows that both types of errors contribute through an OTOC \cite{shenkerBlackHolesButterfly2014,Roberts:2014isa,Maldacena_2016,Shenker:2014cwa,kitaev2015simple}, which grows exponentially in time as $e^{\varkappa t}$, where $\varkappa$ is the quantum Lyapunov exponent. This leads to an exponential deviation $F_{1}(t)\approx 1-\# e^{\varkappa t}$.  

    We proceed to the multi-round Loschmidt echo. A graphical representation, using the two-round case as an example, is shown in FIG.~\ref{fig:contour}(b). For incoherent errors, the insertions of $L_k$ and $L_k^\dagger$ must occur within the same evolution superoperator, either $e^{\mathcal{L}_{H}t}$ or $e^{\mathcal{L}_{-\tilde{H}}t}$. In FIG.~\ref{fig:contour}(b), this corresponds to pairs of branches with the same color, namely $(1,8)$, $(2,7)$, $(3,6)$, and $(4,5)$. Consequently, within perturbation theory, the contribution from incoherent errors is amplified by a factor of $n$. In contrast, two insertions of the coherent error $\delta H$ can occur independently on the backward-evolution branches $2$, $4$, $5$, and $7$. In particular, an OTOC between $\delta H$ and $O$ arises when one $\delta H$ insertion is on branch $2$ or $4$, and the other is on branch $5$ or $7$. Generalizing this discussion to arbitrary $n$ immediately yields an $n^2$ scaling for the accumulation of coherent errors. Putting all ingredients together, the result reads
    \begin{equation}
    \begin{aligned}
    F_n(t) \approx&1- 2n\int_0^t dt'~\sum_k\langle OL_k^\dagger(t') [L_k(t'),O]\rangle\\&-n^2\int_0^t dt'dt''~\frac{1}{2}\langle [O,\delta H(t')][\delta H(t''),O]\rangle.
    \end{aligned}
    \end{equation}
    Therefore, the study of multi-round Loschmidt echoes provides a direct means to distinguish coherent from incoherent errors. In particular, the quadratic scaling of the coherent error arises because, within perturbation theory, inter-round contributions are identical to intra-round ones, reflecting the full coherence of the evolution in perturbation theory.

    Does this distinction persist in the long-time limit? As the evolution time $t$ increases, higher-order corrections must be taken into account. These corrections diminish correlations between inter-round branches and therefore suppress the corresponding $n(n-1)$ contributions that involve inter-round operator pairs. To illustrate this, we analyze the correlation between branches $(2,5)$ by considering the two-point function of $\delta H$ near time $t$, as indicated by the red dots in Fig.~\ref{fig:contour}(b). Similar to previous calculations, this two-point function acquires corrections from both coherent and incoherent errors, which can form an OTOC. An example is an incoherent error occurring at $(2,7)$ near $t \approx 0$, shown as black dots in Fig.~\ref{fig:contour}(b). Consequently, the inter-round correlations are expected to decay with increasing time. In contrast, all contributions from OTOCs cancel in the perturbative expansion of intra-round correlators due to unitarity. As a result, at sufficiently long times, we expect that the accumulation of coherent errors is dominated by intra-round contributions and is therefore expected to scale linearly with the number of rounds, similar to the incoherent case. This physical intuition will be justified by explicit calculations in the next section.

    \emph{ \color{blue}Scramblon Theory.--} To obtain a Loschmidt echo that remains well behaved and saturates to zero at long times, higher-order contributions must be resummed. Scramblon theory provides an efficient framework for this resummation \cite{Kitaev:2017awl,Gu:2018jsv,Zhang:2020jhn,Gu:2021xaj,Stanford:2021bhl,Zhang:2022knu,Zhang:2022fma,Liu:2023lyu,Stanford:2023npy,Zhang:2024vsa,Zhang:2025ckq,Chen:2024imd,Li:2025civ,Perugu:2025vty}. The central assumption is that, for long-time observables with $\varkappa t \gg 1$, OTO correlations dominate the dynamics and are mediated by collective modes known as scramblons. The scramblon propagator takes the form $-\lambda_t\equiv -e^{\varkappa t}/C$, which serves as a signature of quantum many-body chaos. Here, $C \propto N$ and $N$ denotes the number of qubits or fermionic modes. Within scramblon theory, a pair of operators $V$ and $V^\dagger$ in the past can emit scramblons, which are subsequently absorbed by another pair of operators $W$ and $W^\dagger$ in the future, provided that these four operators form an OTOC. The scattering vertex involving $V(t_1)$, $V^\dagger(t_2)$, and $m$ scramblons is denoted by $\Upsilon_V^m(t_{12})$, where we assume time-reversal symmetry \footnote{For systems without time-reversal symmetry, it is necessary to distinguish between scattering vertices located in the past and those in the future by introducing $\Upsilon_V^{R/A,m}(t_{12})$.} and define $t_{12} = t_1 - t_2$. 

\begin{figure}[t]\centering
\begin{tabular}{c@{\hspace{0.5cm}}c}
\begin{tikzpicture}[scale=0.95]
\node[bvertexnormal] (R) at (-0pt,0pt) {\scriptsize$\Upsilon^m_O$};
\node[svertex] (A1) at (30pt,20pt) {};
\node[svertex] (A3) at (30pt,-20pt) {};

\draw[thick] (R) -- ++(135:23pt) node[left]{\scriptsize$O$};
\draw[thick] (R) -- ++(-135:23pt) node[left]{\scriptsize$O$};
\draw[thick] (A1) -- ++(70:10pt) node[above]{\scriptsize$L_k(t')_2$};
\draw[thick] (A1) -- ++(-20:10pt) node[right]{\scriptsize$L_k^\dagger (t')_7$};
\draw[thick] (A3) -- ++(20:10pt) node[right]{\scriptsize$L_{k'}(t'')_3$};
\draw[thick] (A3) -- ++(-70:10pt) node[below]{\scriptsize $L_{k'}^\dagger(t'')_6$};

\draw[wavy] (R) to (A1);
\draw[wavy] (R) to (A3);

\end{tikzpicture}
&
\begin{tikzpicture}[scale=0.95]
\node[bvertexnormal] (R) at (-0pt,0pt) {\scriptsize$\Upsilon^m_O$};
\node[svertex] (A1) at (30pt,20pt) {};
\node[svertex] (A3) at (30pt,-20pt) {};
\node[svertex] (C1) at (0pt,-30pt) {};

\draw[thick] (R) -- ++(135:23pt) node[left]{\scriptsize$O$};
\draw[thick] (R) -- ++(-135:23pt) node[left]{\scriptsize$O$};
\draw[thick] (A1) -- ++(70:10pt) node[above]{\scriptsize$\delta H(t_1)_2$};
\draw[thick] (A1) -- ++(-20:10pt) node[right]{\scriptsize$\delta H (t_2)_7$};
\draw[thick] (A3) -- ++(45:10pt) node[right]{\scriptsize$\delta H(t_3)_2$};
\draw[thick] (A3) -- ++(-45:10pt) node[right]{\scriptsize $\delta H(t_4)_5$};

\draw[thick] (C1) -- ++(145:10pt) node[left]{\scriptsize$\delta H(t_5)_2$};
\draw[thick] (C1) -- ++(235:10pt) node[left]{\scriptsize$\delta H (t_6)_7$};

\draw[wavy] (R) to (A1);
\draw[wavy] (R) to (A3);
\draw[wavy] (C1) to (A3);

\end{tikzpicture}
\vspace{3pt}\\
(a) Incoherent errors& (b) Coherent errors
\end{tabular}
\caption{Typical scramblon diagrams contributing to the Loschmidt echo with $n=2$ for (a) incoherent errors and (b) coherent errors. Wavy lines represent scramblon propagators, while solid lines denote microscopic operators. Subscripts on the error operators indicate their branch indices (see FIG.~\ref{fig:contour}). In panel (a), all scramblons emitted by jump operators are absorbed by the operator $O$. In panel (b), scramblons emitted by inter-round errors with $t_1,t_2,t_3,t_4\approx t$ can also be absorbed by errors with $t_5, t_6\approx0$.
}
\label{fig:Diagram}
\end{figure}
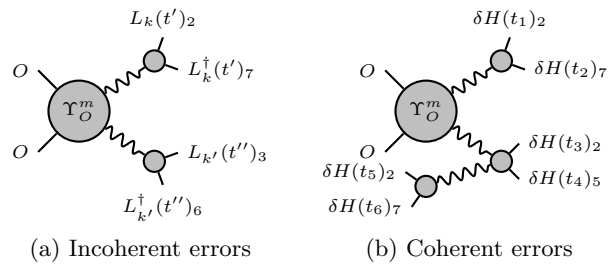

    We first investigate scenarios with either incoherent or coherent errors separately. We begin with the incoherent error case, where each insertion of a jump operator can form an OTOC with the operators $O$, as discussed in the previous section. Within scramblon theory, this amounts to summing diagrams with an arbitrary number of jump-operator insertions $(L_k, L_k^\dagger)$. Each pair of jump operators may originate from an arbitrary Lindbladian superoperator (indicated by different colors in Fig.~\ref{fig:contour}) and can be inserted at arbitrary times. Since the scramblon propagator carries a factor of $1/N$, in the thermodynamic limit $N \to \infty$, each pair of jump operators emits only a single scramblon. All scramblons are subsequently absorbed by the pair of $O$ operators at $t = 0$. An example for $n=2$ is shown in Fig.~\ref{fig:Diagram}(a).

    Summing over all possible scramblon diagrams from incoherent errors, we find
    \begin{equation}\label{eqn_incoherent_dia}
    F_n(t)_I= \sum_{m=0}^\infty\frac{\Upsilon^m_O}{m!}\prod_{j=1}^m\left[\int_0^t dt_j \Big(-2n\lambda_{t_j}\sum_k\Upsilon^1_{L_k}\Big)\right].
    \end{equation}
    For conciseness, we define $\Upsilon^m_V \equiv \Upsilon^m_V(0)$. The subscript indicates only incoherent errors present. The result involves integrations over the insertion times ${t_j}$ of each pair of jump operators, with a total of $m$ insertions. The factor of $2n$ arises from the $n$ copies of forward and backward evolutions. Introducing the auxiliary function $f_V(x)\equiv \sum_m\frac{(-x)^m}{m!}\Upsilon^m_V$, we find 
    \begin{equation}\label{eqn:res1}
    F_n(t)_I=f_O\big(n\gamma_I e^{\varkappa t}\big),\ \ \ \text{with }\gamma_I= \frac{2}{C \varkappa}\sum_k\Upsilon^1_{L_k}.
    \end{equation}
    Here, we assume that $\sum_k \Upsilon^1_{L_k} \propto N$ is extensive in the system size, and consequently $\gamma_I \sim O(1)$. This assumption is valid when the system is coupled homogeneously to the environment. Eq.~\eqref{eqn:res1} clearly shows the linear accumulation of incoherent errors for arbitrary time $t$. For all solvable models in which analytical expressions for $f(x)$ are available, including the large-$q$ SYK model, the Brownian SYK model, and Brownian circuits, we have $f_O(x) = 1/(1+x)^{2\Delta_O}$, with an effective scaling dimension $\Delta_O$ \cite{Gu:2021xaj,Liu:2023lyu}. A similar ansatz has also been demonstrated in solid-state NMR systems \cite{Li:2025civ}. Using this ansatz, the result for incoherent errors is:
    \begin{equation}\label{eqn:resIn}
    F_n(t)_I=\frac{1}{(1+n\gamma_I e^{\varkappa t})^{2\Delta_O}}.
    \end{equation}

    Next, we study the dynamics with coherent errors only. We primarily focus on the $n=2$ case and defer the discussion of general $n$ to the Supplementary Material \cite{SM}. Similar to the incoherent case, the decay of the Loschmidt echo is driven by the exchange of scramblons between a pair of coherent errors $\delta H$ and the operators $O$. As discussed in the previous section, there are four distinct ways of pairing errors on different branches, namely $(2,5)$, $(2,7)$, $(4,5)$, and $(4,7)$, that contribute to this process. However, additional diagrams appear in the coherent error case. As noted earlier, a pair of $\delta H$ operators inserted on branches $(2,5)$ near time $t$ can also form an OTOC with another pair of $\delta H$ operators inserted on branches $(2,7)$ near $t \approx 0$. Consequently, scramblons emitted by $\delta H$ near time $t$ can be absorbed either by the operators $O$ or by the $\delta H$ operators near $t = 0$. A concrete example is shown in Fig.~\ref{fig:Diagram}(b).

    As detailed in the Supplementary Material \cite{SM}, summing the contributions from operators near $t = 0$ effectively renormalizes the scattering vertices for the inter-round terms near time $t$. This leads to
    \begin{equation}\label{eqn:coherent_dia}
    \begin{aligned}
    F_2(t)_c= \sum_{m=0}^\infty\frac{\Upsilon^m_O}{m!}\prod_{j=1}^m\left[\int_0^t dt_j \Big(-2\lambda_{t_j}\big(\bar{\Upsilon}^1_{\delta H}+\tilde{\Upsilon}^1_{\delta H,t_j}\big)\Big)\right].
    \end{aligned}
    \end{equation}
    Here, we have made the Markovian approximation $\Upsilon_{\delta H}^1(t) \approx \delta(t) \bar{\Upsilon}^1_{\delta H}$, which is justified by the fact that the Loschmidt echo decays over a parametrically long timescale in the limit of weak errors. The term $\bar{\Upsilon}^1_{\delta H}$ represents the intra-round perturbations at $(2,7)$ and $(4,5)$, contributing in a manner similar to the incoherent case \eqref{eqn_incoherent_dia}. $\tilde{\Upsilon}^1_{\delta H,t}$ denotes the renormalized scattering vertex, describing additional inter-round contributions that are absent in the incoherent case. The detailed expression for $\tilde{\Upsilon}^1_{\delta H,t}$ is presented in the Supplementary Material \cite{SM}, which leads to
    \begin{equation}\label{eqn:res}
    \begin{aligned}
    F_2(t)_c=f_O\bigg(&2\gamma_ce^{\varkappa t}+\frac{2}{\bar{\Upsilon}^1_{\delta H}}\left[\bar{f}_{\delta H}(0)-\bar{f}_{\delta H}(\gamma_ce^{\varkappa t})\right]\bigg),
    \end{aligned}
    \end{equation}
    where we have introduced the strength of the coherent error as $\gamma_c = \frac{1}{\varkappa C} \bar{\Upsilon}^1_{\delta H}$. Since $\delta H$ is extensive, we expect $\bar{\Upsilon}^1_{\delta H} \propto N$, and therefore $\gamma_c \sim O(1)$ \footnote{ In addition, the normalization of $\gamma_c$ is chosen such that the single-round result satisfies $F_1(t)=f_O(\gamma_c e^{\varkappa t})$, matching the incoherent case.}. Eq.~\eqref{eqn:res} clearly demonstrates the quadratic-to-linear crossover. In the short-time limit, we can expand $\gamma_c e^{\varkappa t}  \ll 1$, giving $F_2(t) \approx f_O\big(4 \gamma_c e^{\varkappa t} \big)$. In contrast, in the late-time regime, the first term in the argument dominates since $\bar{f}_{\delta H}(x)$ is bounded for arbitrary $x>0$ \cite{Gu:2021xaj}. As a consequence, we have $F_2(t) \approx f_O\big(2 \gamma_c e^{\varkappa t} \big)$, which exhibits linear scaling. Closed-form expressions can be obtained by further assuming $\bar{f}_{\delta H}(x) = C_0/(1+bx)^{2\Delta_{\delta}}$, which yields the results for coherent errors
    \begin{equation}\label{eqn:resCo}
    F_2(t)_c=\frac{1}{\big(1+2\gamma_c e^{\varkappa t}+\frac{1}{b\Delta_{\delta }}(1-\frac{1}{(1+b\gamma_c e^{\varkappa t})^{2\Delta_{\delta }}})\big)^{2\Delta_O}}.
    \end{equation}

    Finally, we can combine Eqs.~\eqref{eqn:resIn} and \eqref{eqn:resCo} to obtain the result in the presence of both coherent and incoherent errors for $n=2$. The intra-round contributions simply add, yielding a total error strength $\gamma=\gamma_I+\gamma_c$. These intra-round error pairs also contribute to the renormalization of inter-round coherent error pairs. Putting all these ingredients together, we obtain 
    \begin{equation}\label{eqn:resFull}
    F_2(t)=\frac{1}{\big(1+2\gamma e^{\varkappa t}+\frac{\gamma_c}{b\Delta_{\delta }\gamma}(1-\frac{1}{(1+b\gamma e^{\varkappa t})^{2\Delta_{\delta }}})\big)^{2\Delta_O}}.
    \end{equation}
    This result provides a concrete framework for the characterization and calibration of coherent and incoherent errors in realistic quantum platforms. In particular, single-round time-reversed dynamics have been extensively explored in NMR experiments, for example in measurements of high-order correlations such as multiple quantum coherences \cite{Li:2025civ,PhysRevLett.37.43,10.1063/1.432450,10.1063/1.445185,10.1063/1.449344}. Extending these protocols to multi-round dynamics is straightforward in state-of-the-art NMR systems. We therefore anticipate that our predictions can be readily tested and applied in current experimental setups.

    \emph{ \color{blue}Example: SYK model.--} To further support our theoretical analysis, we provide an explicit verification using the solvable SYK model \cite{Sachdev:1992fk,kitaev2015simple,Maldacena:2016hyu,Kitaev:2017awl,Chowdhury:2021qpy}. The SYK model describes $N$ randomly interacting Majorana fermions $\chi_a$, with $a \in \{1,2,\dots,N\}$. We take the canonical commutation relation $\{\chi_a,\chi_b\}=\delta_{ab}$. The Hamiltonian reads
    \begin{equation}
    H=\sum_{ a<b<c<d}J_{abcd}\chi_{a}\chi_{b}\chi_{c}\chi_{d},
    \end{equation}
    with independent Gaussian variables $\overline{J_{abcd}}=0$ and $\overline{J_{abcd}^2}=3!J^2/N^3$. We consider coherent errors that originate from time-dependent fluctuations of the couplings, taking the form of the Brownian SYK model \cite{P:ramp,Sunderhauf:2019djv}: 
    \begin{equation}
    \delta H(t)=\sum_{ a<b<c<d}V_{abcd}(t)\chi_{a}\chi_{b}\chi_{c}\chi_{d},
    \end{equation}
    where $V_{abcd}(t)$ are Brownian variables with zero expectation and $\overline{V_{abcd}(t)V_{abcd}(t')}=3!V\delta(t-t')/N^3$. To ensure the same structure for coherent and incoherent errors, we choose the jump operator $L_{abcd}=\sqrt{3V/N^3}\chi_a\chi_b\chi_c\chi_d$, identifying $k=abcd$ with $a<b<c<d$. This choice guarantees $\gamma_c=\gamma_I$ by construction. The model with both errors can be analyzed using the large-$N$ expansion, which reduces the quantum dynamics in the thermodynamic limit to self-consistent equations that can be efficiently simulated.

  \begin{figure}[t]
    \centering
    \includegraphics[width=0.75\linewidth]{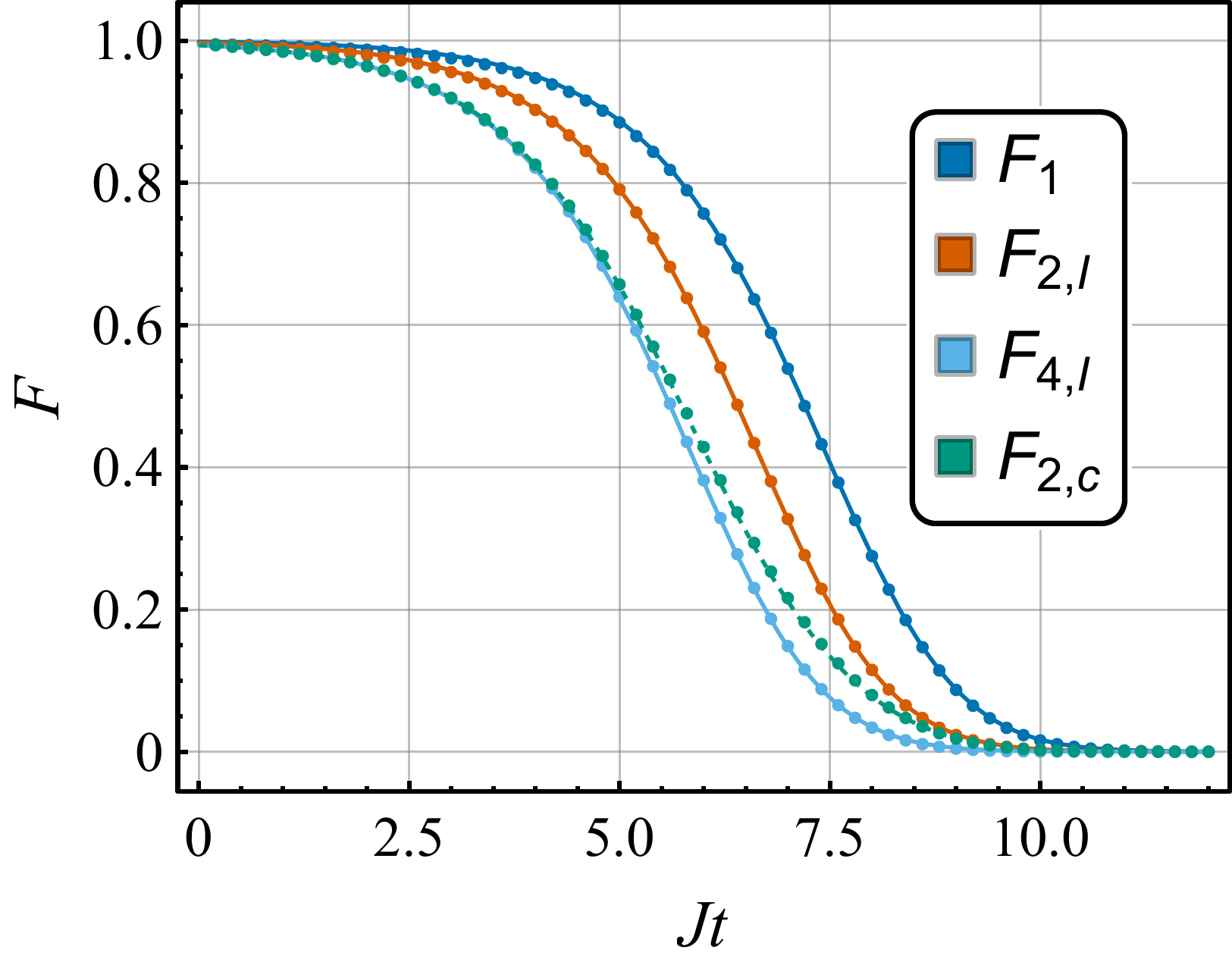}
    \caption{Numerical demonstration of our predictions using the solvable SYK model for the $n$-round Loschmidt echo with either coherent or incoherent errors at $V/J=0.01$. The data points represent results from numerical simulations, while the solid lines are fits based on Eq.~\eqref{eqn:resIn}, and the dashed lines are plotted using Eq.~\eqref{eqn:resCo} with fitted parameters $\gamma_{c}=\gamma_{I}=5.85\times10^{-4}, \varkappa=0.866, \Delta_{O}=1.37$. The results clearly demonstrate a crossover from quadratic to linear scaling for coherent errors.  }
    \label{fig:SYK_num}
  \end{figure}

    Leaving the details to the Supplementary Material \cite{SM}, we present the numerical results in Fig.~\ref{fig:SYK_num} for the operator $O=i\chi_1\chi_2$. We first focus on $F_n(t)_I$ with incoherent errors for $n\in\{1,2,4\}$. The results exhibit the same line shape, with relative time shifts that are approximately equally spaced, consistent with the theoretical prediction \eqref{eqn:resIn}. We further fit the numerical results to \eqref{eqn:resIn}, treating $\gamma_I$, $\varkappa$, and $\Delta_O$ as fitting parameters. The resulting fits, plotted as solid lines in Fig.~\ref{fig:SYK_num}, match the numerical data with high accuracy. Next, we consider the two-round Loschmidt echo $F_2(t)_c$ for systems with coherent errors. The numerical results show that the short-time behavior matches $F_4(t)_I$, while the late-time behavior matches $F_2(t)_I$, clearly demonstrating the quadratic-to-linear crossover. Finally, we plot \eqref{eqn:resCo} in the dashed line using the fitted values $\gamma_c=\gamma_I$, $\varkappa$, and $\Delta_O$, together with exact relations $\Delta_\delta=2\Delta_O$ and $b=1$ (see Supplementary Material \cite{SM}). The resulting curves match the numerical data with good accuracy.

    \emph{ \color{blue}Discussions.--} In this Letter, we investigate the signatures of coherent and incoherent errors in multi-round time-reversed dynamics. We derive concrete results for the Loschmidt echo using scramblon theory \cite{Kitaev:2017awl,Gu:2018jsv,Zhang:2020jhn,Gu:2021xaj,Stanford:2021bhl,Zhang:2022knu,Zhang:2022fma,Liu:2023lyu,Stanford:2023npy,Zhang:2024vsa,Zhang:2025ckq,Chen:2024imd,Li:2025civ,Perugu:2025vty}, which clearly demonstrate distinct scalings for error accumulation. Incoherent errors accumulate linearly with the number of rounds, whereas coherent errors exhibit a crossover from quadratic scaling at short times to linear scaling at late times. We explicitly test these theoretical predictions using a concrete solvable SYK model. Our results can be readily explored in NMR experimental systems.

    We conclude with several remarks. First, although we focus on systems with all-to-all interactions, the perturbative analysis remains valid for arbitrary interacting systems. Therefore, we expect the quadratic-to-linear crossover induced by coherent errors to be a general feature. Second, it may be possible to derive self-consistent relations between Loschmidt echoes with different $n$, providing a way to test scramblon theory without assuming the form of $f_V(x)$ \cite{Liu:2023lyu}. Finally, while we focus on the Loschmidt echo, it would be interesting to generalize our analysis to other observables, such as the fidelity of many-body teleportation \cite{Gao:2016bin,Maldacena:2017axo,Susskind:2017nto,Gao:2018yzk,Brown:2019hmk,Gao:2019nyj,Nezami:2021yaq,PhysRevX.12.031013,Jafferis:2022crx,Liu:2024nhs,Zhou:2024osg,Tian-GangZhou:2024vxm}, in the multi-round setting and investigate their signatures of errors. We leave these directions to future work.

    \vspace{5pt}
    \textit{Acknowledgement.} 
    We thank Hanteng Wang, Zhi-Cheng Yang, Shuo Zhang, Yuke Zhang, and Tian-Gang Zhou for helpful discussions. This project is supported by the NSFC under grant 12374477, the Shanghai Rising-Star Program under grant number 24QA2700300, and the Quantum Science and Technology-National Science and Technology Major Project 2024ZD0300101.

\bibliography{draft.bbl}

\ifarXiv


\section*{Supplementary material}

# Supplementary Material for "Distinguishing Coherent and Incoherent Errors in Multi-Round Time-Reversed Dynamics via Scramblons"

Zeyu Liu[1] and Pengfei Zhang[1,2,*]

[1]*State Key Laboratory of Surface Physics & Department of Physics, Fudan University, Shanghai, 200438, China*
[2]*Hefei National Laboratory, Hefei 230088, China*
(Dated: January 8, 2026)

In this supplementary material, we present: (1) A brief introduction to scramblon theory. (2) Effective renormalization of scattering vertices. (3) Discussion of a general number of rounds $n$. (4) Details of SYK numerics. (5) Exact calculations using scramblon theory verifying the behavior of Loschmidt echo with coherent error.

## I. A BRIEF INTRODUCTION TO SCRAMBLON THEORY

Scramblon theory was proposed as a universal description of information scrambling in quantum chaotic systems [1–4]. The key assumption is that, for weak perturbations, out-of-time-order correlations dominate the dynamics and are mediated by collective modes known as scramblons. Within this framework, a pair of operators $V$ and $V^\dagger$ in the past can emit scramblons, which are subsequently absorbed by another pair of operators $W$ and $W^\dagger$ in the future, provided that these four operators form an OTOC.

The scattering vertex involving the emission from $V(t_2)$, $V^\dagger(t_4)$, and the creation of $m$ scramblons is denoted by $\Upsilon_V^m(t_{24})$, where we assume time-reversal symmetry and define $t_{24} = t_2 - t_4$. For OTOC at infinite temperature, we have

$$F_{W,V} = \pm \langle W(t_1) V(t_2) W^\dagger(t_3) V^\dagger(t_4) \rangle$$
$$= \quad\text{(diagrams)}\quad + \cdots \tag{1}$$
$$= \sum_{m=0}^{\infty} \frac{(-\lambda)^m}{m!} \Upsilon_W^m(t_{13}) \Upsilon_V^m(t_{24}),$$

where the time ordering satisfies $t_1 \approx t_3 \gg t_2 \approx t_4$. In the first line, the sign is $-1$ when $W$ and $V$ are both fermionic operators, and $+1$ otherwise. In the third line, the factor $1/m!$ accounts for the symmetry factor. The scramblon propagator takes the form $\lambda = C^{-1} e^{\varkappa(t_1+t_3-t_2-t_4)/2}$. Here, $C \propto N$ and $N$ denotes the number of qubits or fermionic modes.

We introduce the auxiliary functions

$$f(x,t) = \sum_{m=0}^{\infty} \frac{(-x)^m}{m!} \Upsilon^m(t) = \int_0^\infty dy\, h(y,t) e^{-xy}, \qquad \Upsilon^m(t) = \int_0^\infty dy\, h(y,t) y^m. \tag{2}$$

Substituting these definitions back into $F_{W,V}$, we can resum it into a compact integral form

$$F_{W,V} = \sum_{m=0}^{\infty} \frac{(-\lambda)^m}{m!} \Upsilon_W^m(t_{13}) \Upsilon_V^m(t_{24})$$
$$= \sum_{m=0}^{\infty} \frac{1}{m!} \int_0^\infty dx\, dy\, h_W(x, t_{13}) h_V(y, t_{24}) (-\lambda x y)^m \tag{3}$$
$$= \int_0^\infty dx\, h_W(x, t_{13}) f_V(\lambda x, t_{24}).$$

## II. EFFECTIVE RENORMALIZATION OF SCATTERING VERTICES

Now we turn our attention to the 2-round Loschmidt echo. A graphical representation is shown in Fig. 1. As noted in the main text, a pair of $\delta H$ operators inserted on branches [2, 5] near time $t$ can form an OTOC with pairs of $\delta H$ operators inserted

---

[*] PengfeiZhang.physics@gmail.com

FIG. 1. Graphical representation of the Loschmidt echo $F_n(t)_c$ for $n = 2$. Branches with the same color originate from the same Lindbladian evolution. Solid lines correspond to evolution governed by $\mathcal{L}_H$. Double lines denote evolution in the presence of coherent errors, where insertions of $\delta H$, marked by red dots, may occur. All branches are labeled from 1 to 8.

on branches $[2, 7]$, $[4, 7]$, or $[5, 7]$ near $t \approx 0$. This implies that scramblons emitted by $\delta H$ near time $t$ can be absorbed either by the operators $O$ or by the $\delta H$ operators near $t = 0$. This effect can be captured by defining a renormalized scattering vertex. The diagrammatic expansion for this renormalization is given by

$$
\tag{4}
$$

Here each pair of vertices exchanges a single scramblon in the thermodynamic limit $N \to \infty$ because the scramblon propagator carries a factor of $1/N$. Translating the diagrammatic expansion into an analytical expression involves summing over all possible configurations and integrating over their times. Explicitly, this reads

$$-\lambda_t \Upsilon^1_O \tilde{\Upsilon}^1_{\delta H} = -\lambda_t \Upsilon^1_O \sum_{m_{2,7}=0}^{\infty} \sum_{m_{4,7}=0}^{\infty} \sum_{m_{5,7}=0}^{\infty} \frac{1}{m_{2,7}!} \frac{1}{m_{4,7}!} \frac{1}{m_{5,7}!} (-1)^{m_{5,7}} \prod_{j=1}^{m_{2,7}} \prod_{k=1}^{m_{4,7}} \prod_{l=1}^{m_{5,7}} \int_0^t dt_j dt'_k dt''_l$$
$$\left[-\lambda_{t-t_j}\right]^{m_{2,7}} \left[-\lambda_{t-t'_k}\right]^{m_{4,7}} \left[-\lambda_{t-t''_l}\right]^{m_{5,7}} \left(\tilde{\Upsilon}^1_{\delta H}\right)^{m_{2,7}+m_{4,7}+m_{5,7}} \bar{\Upsilon}^{m_{2,7}+m_{4,7}+m_{5,7}+1}_{\delta H}. \tag{5}$$

Here, the non-negative integers $m_{2,7}$, $m_{4,7}$, and $m_{5,7}$ denote the number of times the corresponding pair of vertices appears in the expansion. We have made the Markovian approximation $\Upsilon^1_{\delta H}(t) \approx \delta(t)\bar{\Upsilon}^1_{\delta H}$, which is justified by the fact that the Loschmidt echo decays over a parametrically long timescale in the weak error limit. The sign factor $(-1)^{m_{5,7}}$ arises from the phase factor in the unitary evolution $\exp(i\delta H t)$. After performing the time integrals and resumming the series, we obtain

$$\tilde{\Upsilon}^1_{\delta H} = \sum_{m_{2,7}=0}^{\infty} \sum_{m_{4,7}=0}^{\infty} \sum_{m_{5,7}=0}^{\infty} \frac{1}{m_{2,7}!} \frac{1}{m_{4,7}!} \frac{(-1)^{m_{5,7}}}{m_{5,7}!} \prod_{j=1}^{m_{2,7}} \prod_{k=1}^{m_{4,7}} \prod_{l=1}^{m_{5,7}} \int_0^t dt_j dt'_k dt''_l$$
$$\left[-\lambda_{t-t_j}\right]^{m_{2,7}} \left[-\lambda_{t-t'_k}\right]^{m_{4,7}} \left[-\lambda_{t-t''_l}\right]^{m_{5,7}} \left(\tilde{\Upsilon}^1_{\delta H}\right)^{m_{2,7}+m_{4,7}+m_{5,7}} \bar{\Upsilon}^{m_{2,7}+m_{4,7}+m_{5,7}+1}_{\delta H}$$
$$= \sum_{m_{2,7}=0}^{\infty} \sum_{m_{4,7}=0}^{\infty} \sum_{m_{5,7}=0}^{\infty} \frac{1}{m_{2,7}!} \frac{1}{m_{4,7}!} \frac{(-1)^{m_{5,7}}}{m_{5,7}!} \left[-\frac{1}{\varkappa}\lambda_t\right]^{m_{2,7}+m_{4,7}+m_{5,7}} \left(\tilde{\Upsilon}^1_{\delta H}\right)^{m_{2,7}+m_{4,7}+m_{5,7}} \int_0^{\infty} dx \, \bar{h}_{\delta H}(x) x^{m_{2,7}+m_{4,7}+m_{5,7}+1} \tag{6}$$
$$= \int_0^{\infty} dx \, \bar{h}_{\delta H}(x) x \exp\left(-\frac{1}{\varkappa}\lambda_t \tilde{\Upsilon}^1_{\delta H} x\right)$$
$$= -\bar{f}'_{\delta H}\left(\gamma_c e^{\varkappa t}\right),$$

where we have introduced the strength of the coherent error as $\gamma_c = \frac{1}{\varkappa C}\bar{\Upsilon}^1_{\delta H}$.

## III. DISCUSSION OF GENERAL NUMBER OF ROUNDS N

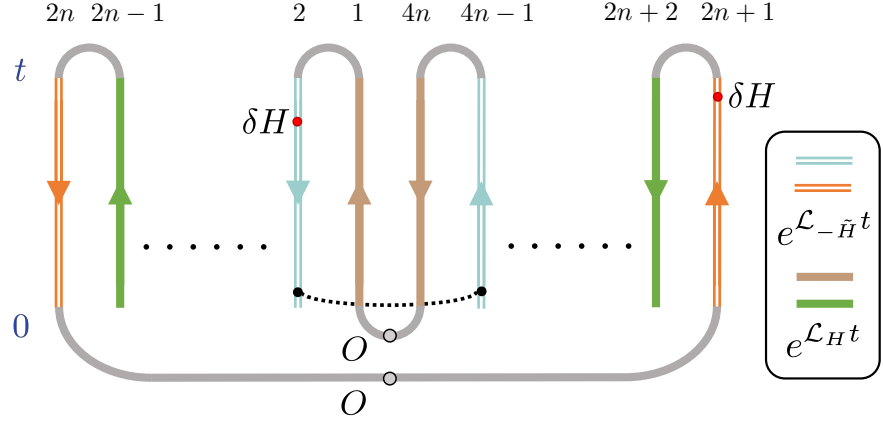

FIG. 2. Graphical representation of the Loschmidt echo $F_n(t)_c$ for general $n$. Branches with the same color originate from the same Lindbladian evolution. Solid lines correspond to evolution governed by $\mathcal{L}_H$, while dotted lines indicate the Lindblad term. Double lines denote evolution in the presence of coherent errors, where insertions of $\delta H$, marked by red dots, may occur. All branches are labeled from 1 to $4n$.

Now we generalize the analysis to the $n$-round Loschmidt echo with both coherent and incoherent errors. A graphical representation is shown in Fig. 2. A pair of $\delta H$ operators inserted on branches $[2p, 2(2n-q)+1]$ near time $t$, where $1 \leq p, q \leq n$, can form an OTOC with pairs of $\delta H$ operators inserted on the following branches

$$
\begin{aligned}
&[2\alpha, 2(2n-\beta)+1], && 1 \leq \alpha < p \text{ and } q \leq \beta \leq n. \\
&[2\alpha, 2(2n-\beta)+1], && p \leq \alpha \leq n \text{ and } 1 \leq \beta < q. \\
&[2\alpha, 2\beta], && 1 \leq \alpha < p \text{ and } p \leq \beta \leq n. \\
&[2(2n-\alpha)+1, 2(2n-\beta)+1], && 1 \leq \alpha < q \text{ and } q \leq \beta \leq n.
\end{aligned}
\tag{7}
$$

For the third and the last cases, there will be a $-1$ sign factor as discussed in the previous section. Additionally, the intra-round error pairs $L_k$ on branch $[m, 4n+1-m]$, with $\min(2p, 2q) \leq m < \max(2p, 2q)$, also contribute to the renormalization of inter-round coherent error pairs. By summing the contributions from all valid branch combinations and incorporating the associated sign factors, the renormalized vertex takes the compact form

$$
\begin{aligned}
\tilde{\Upsilon}_{\delta H} &= -\bar{f}'_{\delta H}\left([(p-1)(n-q+1)+(q-1)(n-p+1)-(p-1)(n-p+1)-(q-1)(n-q+1)]\gamma_c e^{\varkappa t} + |p-q|\gamma_I e^{\varkappa t}\right) \\
&= -\bar{f}'_{\delta H}\left[\left((p-q)^2 \gamma_c + |p-q|\gamma_I\right)e^{\varkappa t}\right].
\end{aligned}
\tag{8}
$$

Here $\gamma_I = \frac{2}{C\varkappa}\sum_k \Upsilon^1_{L_k}$. Using the renormalized vertices, the n-round Loschmidt echo becomes

$$
\begin{aligned}
F_n(t) &= \sum_{m=0}^{\infty} \frac{\Upsilon^m_O}{m!} \prod_{j=1}^{m}\left[\int_0^t dt_j \left(-2n\lambda_{t_j}\sum_k \Upsilon^1_{L_k} + \lambda_{t_j}\sum_{p=1}^{n}\sum_{q=1}^{n}\bar{f}'_{\delta H}\left[\left((p-q)^2\gamma_c + |p-q|\gamma_I\right)e^{\varkappa t_j}\right]\right)\right] \\
&= \sum_{m=0}^{\infty} \frac{\Upsilon^m_O}{m!}\prod_{j=1}^{m}\left[\int_0^t dt_j\left(-2n\lambda_{t_j}\sum_k \Upsilon^1_{L_k} - n\lambda_{t_j}\bar{\Upsilon}^1_{\delta H} + 2\lambda_{t_j}\sum_{a=1}^{n-1}(n-a)\bar{f}'_{\delta H}\left[\left(a^2\gamma_c + a\gamma_I\right)e^{\varkappa t_j}\right]\right)\right] \\
&= \sum_{m=0}^{\infty}\frac{\Upsilon^m_O}{m!}\left(-n\gamma e^{\varkappa t} - 2\sum_{a=1}^{n-1}\frac{n-a}{a^2\tilde{\Upsilon}^1_{\delta H}+2a\sum_k \Upsilon^1_{L_k}}\left[\bar{f}_{\delta H}(0)-\bar{f}_{\delta H}\left((a^2\gamma_c+a\gamma_I)e^{\varkappa t}\right)\right]\right)^m \\
&= f_O\left(n\gamma e^{\varkappa t} + 2\sum_{a=1}^{n-1}\frac{n-a}{a^2\tilde{\Upsilon}^1_{\delta H}+2a\sum_k \Upsilon^1_{L_k}}\left[\bar{f}_{\delta H}(0)-\bar{f}_{\delta H}\left((a^2\gamma_c+a\gamma_I)e^{\varkappa t}\right)\right]\right).
\end{aligned}
\tag{9}
$$

This result is a generalization of Eq. (9) in the main text. In the short-time limit, we can expand for small arguments $\gamma_c e^{\varkappa t} \ll 1$ using Eq. (2), which leads to

$$
F_n(t) = f_O\left((n^2\gamma_c + n\gamma_I)e^{\varkappa t}\right).
\tag{10}
$$

In contrast, For in the late-time limit, the first term in the argument dominates, since $\bar{f}_{\delta H}(x)$ is bounded for arbitrary $x > 0$. As a consequence, we have

$$F_n(t) \approx f_O(n\gamma e^{\varkappa t}), \tag{11}$$

which exhibits linear scaling with $n$.

## IV. DETAILS OF SYK NUMERICS

The SYK model describes $N$ randomly interacting Majorana fermions $\chi_a$, with $a \in \{1, 2, \ldots, N\}$. We adopt the canonical commutation relation $\{\chi_a, \chi_b\} = \delta_{ab}$. The Hamiltonian is given by

$$H = \sum_{a<b<c<d} J_{abcd} \chi_a \chi_b \chi_c \chi_d, \tag{12}$$

with independent Gaussian variables $\overline{J_{abcd}} = 0$ and $\overline{J^2_{abcd}} = 3!J^2/N^3$. We consider coherent errors that originate from time-dependent fluctuations of the couplings, taking the form of the Brownian SYK model:

$$\delta H(t) = \sum_{a<b<c<d} V_{abcd}(t) \chi_a \chi_b \chi_c \chi_d, \tag{13}$$

where $V_{abcd}(t)$ are Brownian variables with zero expectation and $\overline{V_{abcd}(t)V_{abcd}(t')} = 3!V\delta(t-t')/N^3$. To ensure the same structure for coherent and incoherent errors, we choose the jump operator $L_{abcd} = \sqrt{3V/N^3}\chi_a\chi_b\chi_c\chi_d$, identifying $k = abcd$ with $a < b < c < d$. This choice guarantees $\gamma_c = \gamma_I$ by construction.

### A. $n = 1$ with incoherent error

For the 1-round Loschmidt echo with incoherent error, the time domain along the Keldysh contour is denoted as $u \in [0, 4t]$, divided into forward and backward evolution branches

$$\begin{aligned} U_+ &= [0, t] \cup [2t, 3t] \quad \text{forward} \\ U_- &= [t, 2t] \cup [3t, 4t] \quad \text{backward.} \end{aligned} \tag{14}$$

The bare Green's function and the derivative operator on the contour are defined as

$$\begin{aligned} G_0(u, u') &= \frac{1}{2}\text{sgn}(u - u') \\ \partial_u &= G_0^{-1}. \end{aligned} \tag{15}$$

In the large $N$ limit, we have Schwinger-Dyson equations

$$\begin{aligned} G(u, u') &= [\partial_u - \Sigma(u, u')]^{-1} \\ \Sigma(u, u') &= J^2 f(u)f(u')G(u, u')^3 + \frac{V}{2}f(u)f(u')g(u, u')G(u, u')^3, \end{aligned} \tag{16}$$

where the phase factors $f(u)$ distinguish between the forward and backward branches

$$\begin{aligned} f(u) &= i \quad u \in \text{forward} \\ f(u) &= -i \quad u \in \text{backward.} \end{aligned} \tag{17}$$

The function $g$ takes the form

$$g(u, u') = \delta(u - u') + \delta(4t - u - u'). \tag{18}$$

We calculate $(2G(0, 2t))^2$ (as elaborated in Subsection IV D) numerically by solving (16).

The generalization to general $n$ with incoherent errors is straightforward.

## B.  $n = 2$ with coherent error

For the 2-round Loschmidt echo with coherent error, the time domain along the Keldysh contour extends to $u \in [0, 8t]$, where

$$U_+ = [0, t] \cup [2t, 3t] \cup [4t, 5t] \cup [6t, 7t] \quad \text{forward}$$
$$U_- = [t, 2t] \cup [3t, 4t] \cup [5t, 6t] \cup [7t, 8t] \quad \text{backward.} \tag{19}$$

The bare Green's function and the derivative operator on the contour are defined as

$$G_0(u, u') = \frac{1}{2}\text{sgn}(u - u')$$
$$\partial_u = G_0^{-1}. \tag{20}$$

In the large $N$ limit, we have Schwinger-Dyson equations

$$G(u, u') = [\partial_u - \Sigma(u, u')]^{-1}$$
$$\Sigma(u, u') = J^2 f(u) f(u') G(u, u')^3 + V f(u) f(u') g(u, u') G(u, u')^3. \tag{21}$$

The phase factor $f(u)$ follows the same convention as before

$$f(u) = i \quad u \in \text{forward}$$
$$f(u) = -i \quad u \in \text{backward.} \tag{22}$$

For $u \leq u'$ (note that $g$ is symmetric), the function $g$ decomposes into 7 components

$$g(u, u') = \delta(u - u') + \sum_{j=1}^{6} g_j(u, u'), \tag{23}$$

where each term corresponds to a specific pairing of time intervals

$$\begin{aligned}
g_1(u, u') &= \delta(u' - u - 2t) & u &\in [t, 2t] \\
g_2(u, u') &= \delta(u' + u - 6t) & u &\in [t, 2t] \\
g_3(u, u') &= \delta(u' + u - 8t) & u &\in [t, 2t] \\
g_4(u, u') &= \delta(u' + u - 8t) & u &\in [3t, 4t] \\
g_5(u, u') &= \delta(u' + u - 10t) & u &\in [3t, 4t] \\
g_6(u, u') &= \delta(u' - u - 2t) & u &\in [4t, 5t].
\end{aligned} \tag{24}$$

We calculate $(2G(0, 4t))^2$ (as elaborated in Subsection IV D) numerically by solving (21).

## C.  Convolution theorem

This subsection provides the mathematical justification which will be useful in the following subsection. We start by defining the convolution of the auxiliary function with itself

$$h^{(2)}(y, t) = \int_0^y dy' \, h(y - y', t) h(y', t). \tag{25}$$

Applying the Laplace transform converts it into a simple product in the Laplace domain

$$\begin{aligned}
f^{(2)}(x, t) &= \int_0^\infty dy \, h^{(2)}(y, t) e^{-xy} \\
&= \int_0^\infty dy \int_0^y dy' \, h(y - y', t) h(y', t) e^{-xy} \\
&= f(x, t)^2.
\end{aligned} \tag{26}$$

This property generalizes to any power $m$

$$f^{(m)}(x, t) = f(x, t)^m$$
$$h^{(m)}(y, t) = \mathcal{L}^{-1}\{f^{(m)}(x, t)\}(y, t). \tag{27}$$

### D. Fitting the numerical results

For all solvable models in which analytical expressions for $f(x)$ are available, including the large-$q$ SYK model, the Brownian SYK model, and Brownian circuits, we have $f_\chi(x) = 1/(1+x)^{2\Delta}$, with an effective scaling dimension $\Delta$ [3, 5].

To apply this to our specific case, we must relate the composite operator $O = i\chi_1\chi_2$ to the fundamental fermionic modes. On the one hand, treating $O$ as a single bosonic entity forming OTOC with some bosonic operator $V$, we have

$$F_{O,V} = \langle O(t_1)V(t_2)O(t_3)V^\dagger(t_4)\rangle$$
$$= \sum_{m=0}^{\infty} \frac{(-\lambda)^m}{m!} \Upsilon_O^m(t_{13})\Upsilon_V^m(t_{24}) \quad (28)$$
$$= \int_0^\infty dx\, h_O(x, t_{13}) f_V(\lambda x, t_{24}).$$

On the other hand, explicitly decomposing $O$ into its constituent fermions, we have

$$F_{O,V} = \langle i\chi_1(t_1)\chi_2(t_1)V(t_2)i\chi_1(t_3)\chi_2(t_3)V^\dagger(t_4)\rangle$$
$$= \int_0^\infty dy_1 dy_2\, h_\chi(y_1, t_{13})h_\chi(y_2, t_{13})f_V(\lambda(y_1 + y_2), t_{24}). \quad (29)$$

Comparing these two representations implies that

$$h_O(y, t) = \int_0^y dy'\, h_\chi(y - y', t)h_\chi(y', t). \quad (30)$$

Applying the convolution theorem, we find

$$f_O(y, t) = f_\chi(y, t)^2 = \frac{1}{(1+x)^{2\Delta_O}}, \quad (31)$$

where the scaling dimension doubles: $\Delta_O = 2\Delta$. Similarly, for the perturbation operator $\delta H$, we have

$$f_{\delta H}(y, t) = C_0 f_\chi(y, t)^4 = \frac{C_0}{(1+bx)^{2\Delta_\delta}}, \quad (32)$$

with $b = 1$ and $\Delta_\delta = 2\Delta_O = 4\Delta$.

Substituting these explicit forms into (9), we obtain

$$F_n(t)_I = \frac{1}{(1 + n\gamma_I e^{\varkappa t})^{2\Delta_O}}, \quad (33)$$

and

$$F_2(t)_c = \frac{1}{\left(1 + 2\gamma_c e^{\varkappa t} + \frac{1}{2\Delta_O}(1 - \frac{1}{(1+\gamma_c e^{\varkappa t})^{4\Delta_O}})\right)^{2\Delta_O}}. \quad (34)$$

To verify these predictions, we compare them with numerical results for the $n$-round Loschmidt echo defined on a time domain $u \in [0, T]$. In terms of correlation function, we have $F_n^\chi(t) = 2G(0, T/2)$, where the factor 2 arises from our normalization convention. Using the convolution property established above, we have

$$F_n(t) \equiv F_n^O(t) = \left(F_n^\chi(t)\right)^2 = (2G(0, T/2))^2. \quad (35)$$

We present the numerical results in Fig. 4 in the main text for the operator.

## V. EXACT CALCULATIONS USING SCRAMBLON THEORY

While our discussion in the previous sections and in the main text provided an intuitive understanding of the Loschmidt echo's behavior, especially the short-time quadratic and late-time linear scaling for coherent errors, a full analytical calculation is essential for a complete verification. In this section, we focus on two simplified models where we can calculate the Loschmidt echo analytically in the thermodynamic limit $N \to \infty$, thereby solidifying our predictions.

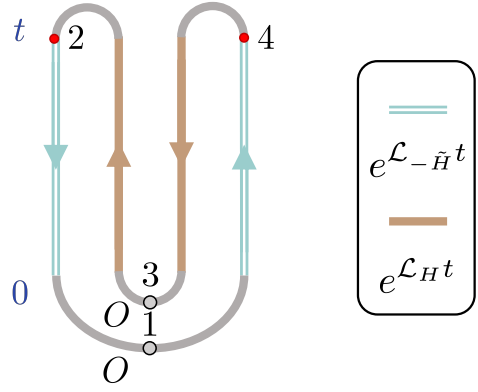

FIG. 3. Graphical representation of the Loschmidt echo $F_1(t)$. Branches with the same color originate from the same Lindbladian evolution. Solid lines correspond to evolution governed by $\mathcal{L}_H$. Double lines denote evolution in the presence of coherent errors, where $\delta H$ exists only at the points labeled by red dots. We label the different operators from 1 to 4.

### A. Model 1

For simplicity, we focus on a 1-round Loschmidt echo where the perturbation $\delta H$ is applied only at time $t$. A graphical representation is shown in Fig. 3. Calculations for this model have been discussed in [6]. For completeness, we present its derivation here. The Loschmidt echo becomes

$$F_1(t) = \langle O(0)_1 e^{i\delta H(t)_2} O(0)_3 e^{-i\delta H(t)_4} \rangle = \sum_{l=0}^{\infty} \sum_{r=0}^{\infty} \frac{1}{l! r!} (i)^l (-i)^r \langle O(0)_1 [\delta H(t)_2]^l O(0)_3 [\delta H(t)_4]^r \rangle. \tag{36}$$

The operators are paired using Wick's theorem to form the OTOC. Here we suppose that $m_2$ pairs of $\delta H(t)_2$ operators are self-contracted, $m_4$ pairs of $\delta H(t)_4$ operators are self-contracted, while $m_{2,4}$ pairs of $\delta H(t)_2$ and $\delta H(t)_4$ are contracted with each other. Then we have

$$l = 2m_2 + m_{2,4}, \qquad r = 2m_4 + m_{2,4}. \tag{37}$$

Among all the $l$ operators in $\delta H(t)_2$, we need to choose $2m_2$ of them for self-contraction, which introduces the binomial coefficient $C_l^{2m_2}$. Similarly, we have the coefficient $C_r^{2m_4}$ for $\delta H(t)_4$. Furthermore, there are $(2m_2 - 1)!!$ and $(2m_4 - 1)!!$ ways to pair the self-contracted operators. We also obtain a factor of $m_{2,4}!$ when applying Wick's theorem to the contractions between $\delta H(t)_2$ and $\delta H(t)_4$. Finally, the prefactor in the summation reads

$$\frac{(i)^{2m_2+m_{2,4}}}{(2m_2+m_{2,4})!} \frac{(-i)^{2m_4+m_{2,4}}}{(2m_4+m_{2,4})!} C_{2m_2+m_{2,4}}^{2m_2} C_{2m_4+m_{2,4}}^{2m_4} m_{2,4}! (2m_2-1)!!(2m_4-1)!!$$
$$= \frac{1}{(m_2)!(m_4)!(m_{2,4})!} \left(-\frac{1}{2}\right)^{m_2+m_4}. \tag{38}$$

One can see that only the pairing of $\delta H(t)_2$ and $\delta H(t)_4$ generates the OTOC structure, which signifies the existence of only a single type of scramblon.

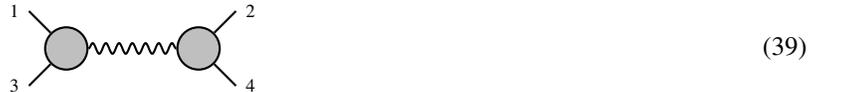
(39)

By summing the contributions from all valid configurations and incorporating the associated sign factors, the Loschmidt echo becomes

$$F_1(t) = \sum_{m_2=0}^{\infty} \sum_{m_4=0}^{\infty} \sum_{m_{2,4}=0}^{\infty} \frac{1}{(m_2)!(m_4)!(m_{2,4})!} \left(-\frac{1}{2}\right)^{m_2+m_4} [G_{\delta H}]^{m_2+m_4} \int_0^{\infty} dx dy_1 \cdots dy_{m_{2,4}}$$
$$h_O(x) h_{\delta H}(y_1) \cdots h_{\delta H}(y_{m_{2,4}}) \exp\left[-\lambda x (y_1 + \cdots + y_{m_{2,4}})\right] \tag{40}$$
$$= \int_0^{\infty} dx\, h_O(x)\, \exp\left[f_{\delta H}(\lambda x) - G_{\delta H}\right].$$

Here $G_{\delta H} \equiv \Upsilon^0_{\delta H}$ is the two-point function. As mentioned earlier, $C$ is proportional to $N$. Hence in the thermodynamic limit $N \to \infty$, we have

$$F_1(t) = \int_0^\infty dx\, h_O(x) \, \exp\left[f_{\delta H}(\lambda x) - G_{\delta H}\right]$$
$$= \int_0^\infty dx\, h_O(x) \, \exp\left[G_{\delta H} - \lambda x \Upsilon^1_{\delta H} - G_{\delta H}\right] \quad (41)$$
$$= f_O\left(\lambda \Upsilon^1_{\delta H}\right).$$

This exact analytical calculation for Model 1, a simplified scenario where the perturbation $\delta H$ acts only at two points, confirms the general form of the Loschmidt echo given by Eq. (9) in the thermodynamic limit, thus solidifying our predictions.

### B. Model 2

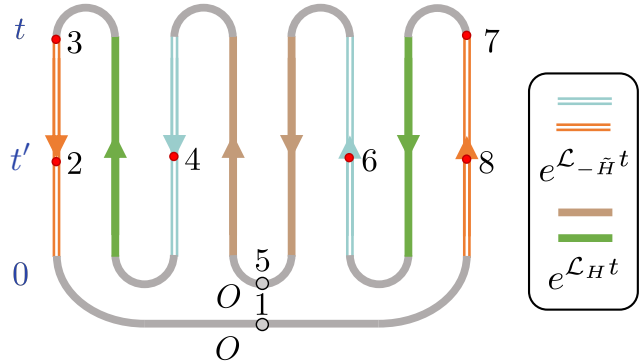

FIG. 4. Graphical representation of the Loschmidt echo $F_2(t)$. Branches of the same color correspond to the same Lindbladian evolution. Solid lines correspond to evolution governed by $\mathcal{L}_H$. Double lines denote evolution in the presence of coherent errors, where $\delta H$ exists only at the points marked by red dots. We label the different operators from 1 to 8.

Now we turn to a 2-round Loschmidt echo where the perturbation $\delta H$ is applied only at time $t$ and $t'$. A graphical representation is shown in Fig. 4.

We find there are 9 types of scramblons

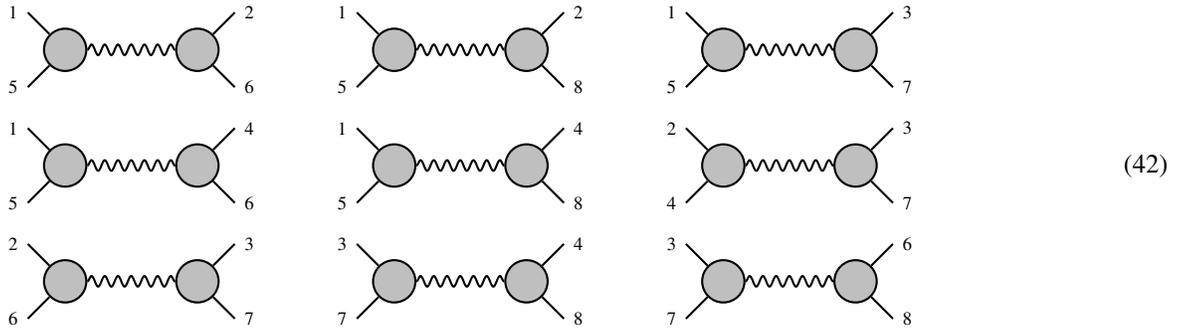

(42)

To systematically account for all possible contraction patterns among the eight operators using Wick's theorem, we need to carefully determine the combinatorial prefactor. This prefactor includes contributions from various sources:

1. Taylor expansion coefficients: These arise directly from the exponential expansion of $e^{i\delta H}$ and $e^{-i\delta H}$. For each operator $\delta H(t)_k$ (where $k \in \{2, 3, 4, 6, 7, 8\}$) appearing $L_k$ times, this introduces a factor of $\frac{(i)^{L_k}}{L_k!}$ or $\frac{(-i)^{L_k}}{L_k!}$ depending on the sign in the exponential. The total contribution is:

$$\mathcal{D}_1 = \frac{(i)^{L_2}}{L_2!} \frac{(i)^{L_3}}{L_3!} \frac{(i)^{L_4}}{L_4!} \frac{(-i)^{L_6}}{L_6!} \frac{(-i)^{L_7}}{L_7!} \frac{(-i)^{L_8}}{L_8!}. \quad (43)$$

2. Arrangements of operators for contraction (derived from permutations): After Taylor expansion, we have $L_k$ identical operators of type $k$. When applying Wick's theorem, these $L_k$ operators are partitioned into groups for self-contractions ($2m_k$ operators) and cross-contractions ($m_{k,j}$ operators for each $j$). The number of ways to arrange these operators for specific contraction patterns is given by a multinomial coefficient for each operator type:

$$\mathcal{D}_2 = \frac{L_2!}{(2m_2)!(m_{2,4})!(m_{2,6})!(m_{2,8})!} \cdot \frac{L_3!}{(2m_3)!(m_{3,7})!} \cdot \frac{L_4!}{(2m_4)!(m_{2,4})!(m_{4,6})!(m_{4,8})!} \\ \cdot \frac{L_6!}{(2m_6)!(m_{2,6})!(m_{4,6})!(m_{6,8})!} \cdot \frac{L_7!}{(2m_7)!(m_{3,7})!} \cdot \frac{L_8!}{(2m_8)!(m_{2,8})!(m_{4,8})!(m_{6,8})!}. \tag{44}$$

3. Ways to form self-contractions (with same index): For each type $k$, if there are $2m_k$ operators that self-contract, the number of ways to pair them up is $(2m_k - 1)!!$. This product over all $k$ is:

$$\mathcal{D}_3 = (2m_2 - 1)!!(2m_3 - 1)!!(2m_4 - 1)!!(2m_6 - 1)!!(2m_7 - 1)!!(2m_8 - 1)!!. \tag{45}$$

4. Ways to form cross-contractions (between different indices): For each pair of operator types $(k, j)$ involved in a cross-contraction, if $m_{k,j}$ pairs are formed, there are $m_{k,j}!$ ways to make these specific pairings. The product over all distinct cross-contraction types is:

$$\mathcal{D}_4 = m_{2,4}!m_{2,6}!m_{2,8}!m_{3,7}!m_{4,6}!m_{4,8}!m_{6,8}!. \tag{46}$$

Finally, the prefactor reads

$$\mathcal{D} = \mathcal{D}_1 \mathcal{D}_2 \mathcal{D}_3 \mathcal{D}_4 \\ = \frac{1}{m_2!m_3!m_4!m_6!m_7!m_8!m_{2,4}!m_{2,6}!m_{2,8}!m_{3,7}!m_{4,6}!m_{4,8}!m_{6,8}!} \left(-\frac{1}{2}\right)^{m_2+m_3+m_4+m_6+m_7+m_8} (-1)^{m_{2,4}+m_{6,8}}. \tag{47}$$

With the combinatorial prefactor now fully determined, we proceed to sum over all possible configurations of $m$ values. By summing the contributions from all valid configurations and incorporating the associated sign factors, the 2-round Loschmidt echo becomes

$$F_2(t) = \sum_{m_2=0}^\infty \sum_{m_3=0}^\infty \sum_{m_4=0}^\infty \sum_{m_6=0}^\infty \sum_{m_7=0}^\infty \sum_{m_8=0}^\infty \sum_{m_{2,4}=0}^\infty \sum_{m_{2,6}=0}^\infty \sum_{m_{2,8}=0}^\infty \sum_{m_{3,7}=0}^\infty \sum_{m_{4,6}=0}^\infty \sum_{m_{4,8}=0}^\infty \sum_{m_{6,8}=0}^\infty \frac{1}{m_2!m_3!m_4!m_6!m_7!m_8!m_{2,4}!m_{2,6}!m_{2,8}!m_{3,7}!m_{4,6}!m_{4,8}!m_{6,8}!} \\ \left(-\frac{1}{2}\right)^{m_2+m_3+m_4+m_6+m_7+m_8} (-1)^{m_{2,4}+m_{6,8}} (G_{\delta H})^{m_2+m_3+m_4+m_6+m_7+m_8} \int_0^\infty dx\, h_O(x) \prod_{i_{2,4}=1}^{m_{2,4}} \prod_{i_{2,6}=1}^{m_{2,6}} \prod_{i_{2,8}=1}^{m_{2,8}} \prod_{i_{3,7}=1}^{m_{3,7}} \prod_{i_{4,6}=1}^{m_{4,6}} \prod_{i_{4,8}=1}^{m_{4,8}} \prod_{i_{6,8}=1}^{m_{6,8}} \bigg( \\ \int_0^\infty dy^{(2,4)}_{i_{2,4}} dy^{(2,6)}_{i_{2,6}} dy^{(2,8)}_{i_{2,8}} dy^{(3,7)}_{i_{3,7}} dy^{(4,6)}_{i_{4,6}} dy^{(4,8)}_{i_{4,8}} dy^{(6,8)}_{i_{6,8}}\, h_{\delta H}\left(y^{(2,4)}_{i_{2,4}}\right) h_{\delta H}\left(y^{(2,6)}_{i_{2,6}}\right) h_{\delta H}\left(y^{(2,8)}_{i_{2,8}}\right) h_{\delta H}\left(y^{(3,7)}_{i_{3,7}}\right) h_{\delta H}\left(y^{(4,6)}_{i_{4,6}}\right) h_{\delta H}\left(y^{(4,8)}_{i_{4,8}}\right) h_{\delta H}\left(y^{(6,8)}_{i_{6,8}}\right) \\ \exp\left\{-\lambda_{t'} x \left(y^{(2,6)}_{i_{2,6}} + y^{(2,8)}_{i_{2,8}} + y^{(4,6)}_{i_{4,6}} + y^{(4,8)}_{i_{4,8}}\right) - \lambda_t x y^{(3,7)}_{i_{3,7}} - \lambda_{t-t'} y^{(3,7)}_{i_{3,7}} \left(y^{(2,4)}_{i_{2,4}} + y^{(2,6)}_{i_{2,6}} + y^{(4,8)}_{i_{4,8}} + y^{(6,8)}_{i_{6,8}}\right)\right\}\bigg) \\ = \sum_{m_{2,4}=0}^\infty \sum_{m_{2,6}=0}^\infty \sum_{m_{2,8}=0}^\infty \sum_{m_{3,7}=0}^\infty \sum_{m_{4,6}=0}^\infty \sum_{m_{4,8}=0}^\infty \sum_{m_{6,8}=0}^\infty \frac{(-1)^{m_{2,4}+m_{6,8}}}{m_{2,4}!m_{2,6}!m_{2,8}!m_{3,7}!m_{4,6}!m_{4,8}!m_{6,8}!} \int_0^\infty dx\, h_O(x) \left[\prod_{i=1}^{m_{3,7}} \int_0^\infty dy^{(3,7)}_i\, h_{\delta H}\left(y^{(3,7)}_i\right)\right] \\ f_{\delta H}[\lambda_{t'} x]^{m_{2,8}+m_{4,6}} f_{\delta H}\left[\lambda_{t-t'}\left(y^{(3,7)}_1 + \cdots + y^{(3,7)}_{m_{3,7}}\right)\right]^{m_{2,4}+m_{6,8}} f_{\delta H}\left[\lambda_{t-t'}\left(y^{(3,7)}_1 + \cdots + y^{(3,7)}_{m_{3,7}}\right) + \lambda_{t'} x\right]^{m_{2,6}+m_{4,8}} \\ \exp\left\{-\lambda_t x\left[y^{(3,7)}_1 + \cdots + y^{(3,7)}_{m_{3,7}}\right] - 3G_{\delta H}\right\} \\ = \sum_{m=0}^\infty \frac{1}{m!} \int_0^\infty dx\, h_O(x) \left[\prod_{i=1}^m \int_0^\infty dy^{(3,7)}_i\, h_{\delta H}\left(y^{(3,7)}_i\right)\right] \exp\bigg\{-\lambda_t x\left[y^{(3,7)}_1 + \cdots + y^{(3,7)}_{m_{3,7}}\right] - 3G_{\delta H} + 2f_{\delta H}[\lambda_{t'} x] \\ + 2f_{\delta H}\left[\lambda_{t-t'}\left(y^{(3,7)}_1 + \cdots + y^{(3,7)}_{m_{3,7}}\right) + \lambda_{t'} x\right] - 2f_{\delta H}\left[\lambda_{t-t'}\left(y^{(3,7)}_1 + \cdots + y^{(3,7)}_{m_{3,7}}\right)\right]\bigg\}. \tag{48}$$

Applying the convolution theorem, we find

$$
\begin{aligned}
F_2(t) &= \sum_{m=0}^{\infty} \frac{1}{m!} \int_0^{\infty} dx\, h_O(x) \int_0^{\infty} dy\, h_{\delta H}^{(m)}(y) \exp\left\{-\lambda_t xy - 3G_{\delta H} + 2f_{\delta H}[\lambda_{t'} x] + 2f_{\delta H}[\lambda_{t-t'} y + \lambda_{t'} x] - 2f_{\delta H}[\lambda_{t-t'} y]\right\} \\
&= \sum_{m=0}^{\infty} \frac{1}{m!} \int_0^{\infty} dx\, h_O(x) \int_0^{\infty} dy\, \mathcal{L}^{-1}\{f_{\delta H}(z)^m\}(y) \exp\left\{-\lambda_t xy - 3G_{\delta H} + 2f_{\delta H}[\lambda_{t'} x] + 2f_{\delta H}[\lambda_{t-t'} y + \lambda_{t'} x] - 2f_{\delta H}[\lambda_{t-t'} y]\right\} \\
&= \int_0^{\infty} dx\, h_O(x) \int_0^{\infty} dy\, \mathcal{L}^{-1}\{\exp[f_{\delta H}(z)]\}(y) \exp\left\{-\lambda_t xy - 3G_{\delta H} + 2f_{\delta H}[\lambda_{t'} x] + 2f_{\delta H}[\lambda_{t-t'} y + \lambda_{t'} x] - 2f_{\delta H}[\lambda_{t-t'} y]\right\}.
\end{aligned}
\tag{49}
$$

Now, to obtain a concrete analytical result, we consider a specific functional form for $f_{\delta H}(z)$. Suppose we have

$$
f_{\delta H}[z] = \frac{N}{1+z},
\tag{50}
$$

which is the $f$ function for Brownian circuit. Here we add a factor $N$ since we have assumed that $\Upsilon_{\delta H}^m$ is proportional to $N$. With this specific $f_{\delta H}(z)$, we can analyze the integral more closely. We focus on the function

$$
\begin{aligned}
K(y) &= \mathcal{L}^{-1}\{\exp(f_{\delta H}(z) - G_{\delta H})\}(y) \\
&= \mathcal{L}^{-1}\left\{\exp\left(\frac{N}{1+z} - N\right)\right\}(y) \\
&= e^{-N-y}\left(\sqrt{\frac{N}{y}} \mathcal{I}_1(2\sqrt{Ny}) + \delta(y)\right).
\end{aligned}
\tag{51}
$$

Here $\mathcal{I}_1$ is the modified Bessel function. For $y > 0$ in the large $N$ limit, we can omit the second term. It is easy to show

$$
\int_0^{\infty} dy\, K(y) = 1.
\tag{52}
$$

For $z \to \infty$, we have

$$
\mathcal{I}_1(z) = \frac{e^z}{\sqrt{2\pi z}}.
\tag{53}
$$

Applying this approximation to $K(y)$ for large argument $2\sqrt{Ny}$ yields

$$
K(y) = e^{-N-y}\sqrt{\frac{N}{y}} \frac{\exp(2\sqrt{Ny})}{2\sqrt{\pi}(Ny)^{1/4}}.
\tag{54}
$$

To find the dominant contribution of this function, particularly in the large $N$ limit, we employ the saddle-point approximation. We define the exponent as $\phi(y)$

$$
\phi(y) = -N - y + 2\sqrt{Ny}.
\tag{55}
$$

The saddle point is found by setting $\phi'(y) = 0$, which yields $y = N$. Near the saddle point, we have

$$
K(y) = \frac{1}{2\sqrt{\pi N}} \exp\left(-\frac{(y-N)^2}{4N}\right).
\tag{56}
$$

This Gaussian distribution indicates that in the strict large $N$ limit, the function $K(y)$ becomes sharply peaked around $y = N$. Thus, we can approximate it with a Dirac delta function

$$
K(y) = \delta(y - N).
\tag{57}
$$

Generally, we focus on

$$
\begin{aligned}
&\mathcal{L}^{-1}\{\exp[f_{\delta H}(z) - G_{\delta H}]\}(y) \\
&= \frac{1}{2\pi i} \lim_{T \to \infty} \int_{\gamma - iT}^{\gamma + iT} \exp\{[zy + f_{\delta H}(z) - G_{\delta H}]\} dz.
\end{aligned}
\tag{58}
$$

We expect the saddle point of y is given by

$$y = \Upsilon^1_{\delta H}, \tag{59}$$

which gives

$$\mathcal{L}^{-1}\{\exp[f_{\delta H}(z) - G_{\delta H}]\}(y) = \delta(y - \Upsilon^1_{\delta H}). \tag{60}$$

Putting all these ingradients together, we have

$$\begin{aligned}
F_2(t) &= \int_0^\infty dx\, h_O(x) \int_0^\infty dy\, \delta(y - \Upsilon^1_{\delta H})\, \exp\left\{-\lambda_t xy - 2G_{\delta H} + 2f_{\delta H}[\lambda_{t'} x] + 2f_{\delta H}[\lambda_{t-t'} y + \lambda_{t'} x] - 2f_{\delta H}[\lambda_{t-t'} y]\right\} \\
&= \int_0^\infty dx\, h_O(x)\, \exp\left\{-\lambda_t x \Upsilon^1_{\delta H} - 2G_{\delta H} + 2f_{\delta H}[\lambda_{t'} x] + 2f_{\delta H}[\lambda_{t-t'} \Upsilon^1_{\delta H} + \lambda_{t'} x] - 2f_{\delta H}[\lambda_{t-t'} \Upsilon^1_{\delta H}]\right\}.
\end{aligned} \tag{61}$$

As mentioned earlier, $C$ is proportional to $N$. Hence we have $\lambda \propto N^{-1}$ and $\Upsilon^1_{\delta H} \propto N$. Expand to order $N^{-1}$, we have

$$\begin{aligned}
F_2(t) &= \int_0^\infty dx\, h_O(x)\, \exp\left\{-\lambda_t x \Upsilon^1_{\delta H} - 2G_{\delta H} + 2G_{\delta H} - 2\lambda_{t'} x \Upsilon^1_{\delta H} + 2\lambda_{t'} x f'_{\delta H}[\lambda_{t-t'} \Upsilon^1_{\delta H}]\right\} \\
&= f_O\left(\lambda_t \Upsilon^1_{\delta H} + 2\lambda_{t'} \Upsilon^1_{\delta H} - 2\lambda_{t'} f'_{\delta H}[\lambda_{t-t'} \Upsilon^1_{\delta H}]\right) \\
&= f_O\left(\lambda_t \Upsilon^1_{\delta H} + 2\lambda_{t'} \Upsilon^1_{\delta H} + 2\lambda_{t'} \tilde{\Upsilon}^1_{\delta H}\right),
\end{aligned} \tag{62}$$

where $\tilde{\Upsilon}^1_{\delta H} \equiv -f'_{\delta H}[\lambda_{t-t'} \Upsilon^1_{\delta H}]$.

This exact analytical calculation for Model 2, a simplified scenario where the perturbation $\delta H$ acts only at six points, further confirms the general form of the Loschmidt echo given by Eq. (9) in the thermodynamic limit, thus solidifying our predictions.

---

[1] A. Kitaev and S. J. Suh, The soft mode in the Sachdev-Ye-Kitaev model and its gravity dual, JHEP **05**, 183, arXiv:1711.08467 [hep-th].
[2] Y. Gu and A. Kitaev, On the relation between the magnitude and exponent of OTOCs, JHEP **02**, 075, arXiv:1812.00120 [hep-th].
[3] Y. Gu, A. Kitaev, and P. Zhang, A two-way approach to out-of-time-order correlators, JHEP **03**, 133, arXiv:2111.12007 [hep-th].
[4] D. Stanford, Z. Yang, and S. Yao, Subleading Weingartens, JHEP **02**, 200, arXiv:2107.10252 [hep-th].
[5] Z. Liu and P. Zhang, Signature of Scramblon Effective Field Theory in Random Spin Models, Phys. Rev. Lett. **132**, 060201 (2024), arXiv:2306.05678 [quant-ph].
[6] Y.-C. Li *et al.*, Error-resilient Reversal of Quantum Chaotic Dynamics Enabled by Scramblons, (2025), arXiv:2506.19915 [cond-mat.str-el].


\fi
	
\end{document}